\documentclass[preprint,aps]{revtex4}
\usepackage{amssymb}
\usepackage{amsmath}

\newcommand{\be}{\begin{equation}}
\newcommand{\ee}{\end{equation}}
\newcommand{\ba}{\begin{eqnarray}}
\newcommand{\ea}{\end{eqnarray}}

\newcommand{\f}{\frac}

\begin{document}

\title{Time-dependent 3D oscillator with Coulomb interaction: an alternative
approach for analyzing quark-antiquark systems}
\author{Jeong Ryeol Choi$^1$, Salim Medjber$^2$, Salah Menouar$^3$, and
Ramazan Sever$^4$ \vspace{0.0cm}}
\affiliation{$^1$School of Electronic Engineering, Kyonggi University, Suwon 16227, Republic of Korea
\vspace{0.0cm}}
\affiliation{$^2$Laboratory of Physical and Chemistry Materials, Faculty of Science,
University of M'sila, M'sila 28000, Algeria \vspace{0.0cm}}
\affiliation{$^3$Department of Physics, Setif1 University-Ferhat Abbas, Setif 19000,
Algeria \vspace{0.0cm}}
\affiliation{$^4$Department of Physics, Middle East Technical University, Ankara 06531,
Turkey \vspace{0.0cm}}

\begin{abstract}
\indent
In this work, the dynamics of quark-antiquark pair systems is investigated
by modelling them as general time-dependent 3D oscillators
perturbed by a Coulomb potential.
Solving this model enables the prediction of key mesonic properties such as
the probability density, energy spectra, and quadrature uncertainties,
offering theoretical insights into
the confinement of quarks via gluon-mediated strong interactions.
To tackle the mathematical
difficulty raised by the time dependence of parameters in the system,
special mathematical techniques, such as the invariant operator method,
unitary transformation method, and the Nikiforov-Uvarov functional analysis
(NUFA) are used.
The wave functions of the system,
derived using these mathematical techniques, are expressed analytically
in terms of the Gauss
hypergeometric function whose mathematical properties are well characterized.
Our results provide
the quantum mechanical framework of quark-antiquark systems
which are essential for exploring
the non-perturbative aspects of QCD.
In addition, the underlying mathematical structure
may serve as a foundation for addressing broader
challenges in particle physics, including
the origin of mass and its connection to the Higgs mechanism.
{\vspace{0.3cm}} \newline
\textbf{Keywords:}{\normalsize \ quark-antiquark systems, wave functions, Nikiforov-Uvarov functional analysis,
Gauss hypergeometric function, invariant operator }

\end{abstract}

\maketitle

{\ \ \ } \newline
\textbf{1. Introduction} \newline
Quark-antiquark pairs are a key concept in particle physics, essential for
understanding the strong interaction and the structure of hadrons such as
mesons and baryons. Typical examples of quark-antiquark pairs are bound
states of a heavy quark and its antiquark, such as charmonium and
bottomonium, collectively referred to as quarkonia \cite{cca}. Quarkonia can be readily
produced in experiments such as the LHC, and their properties can be
measured with high precision.
Quarkonia's distinct mass spectra, unique decay patterns, and relatively
long lifetimes are of significant interest in the search for physics that
extends past the Standard Model. As such, they may offer a promising avenue
for probing potential new phenomena not accounted for by the Higgs sector
\cite{ne1,ne2}. This makes quarkonia an important topic requiring
collaborative research between experimental and theoretical physicists.
Indeed, they are actively studied in connection with new particle searches
and dark matter theories \cite{np1,np2,np3,np4,np5}.

In light of this, it is essential to establish a theoretical framework that
enables systematic analysis and interpretation of quarkonium systems. In
particle physics, quark-antiquark systems are commonly modeled using a
potential that combines harmonic and Coulomb terms \cite{1-,1-1,1-2}, where
the harmonic component captures the long-range confinement effect and the
Coulomb term accounts for the short-range 
interactions based on quantum
chromodynamics (QCD). Motivated by this, we focus on the analysis of
time-dependent 3D oscillators perturbed by a Coulomb potential throughout
this work. The quantum mechanical solutions of such systems are known to be
expressible in terms of biconfluent Heun functions \cite{sun,sun1}. However,
limited knowledge of the properties of such functions continues to make
these systems difficult to analyze. At present, the characterization of these
functions remains confined to their behavior in certain special cases
and their connections to specific recursion
relations, particularly their representations in terms of these
relations \cite{lh1,lh2}. This
lack of comprehensive understanding is the reason why many researchers refer
to them primarily through their defining equation, the biconfluent Heun
equation, rather than treating them as fully characterized functions in
their own right \cite{bch1,bch2,bch3,bch4,bch5}.

To overcome this limitation, the present work explores an alternative
analytical treatment
by expressing the solutions in terms of well-established mathematical
functions, thereby providing a more accessible and rigorous theoretical
understanding of the system. Deriving such solutions to fundamental
equations for our general physical systems governed by time-dependent
Hamiltonians \cite{td1,td2,td3} may require special mathematical techniques beyond conventional
methods such as separation of variables.
One promising approach is the invariant operator method \cite{12,13,30,15}, which
has proven effective in the study of time-dependent Hamiltonian systems
(TDHSs). This method allows for the construction of exact quantum solutions
in systems with time-dependent parameters. Specifically, if an invariant
operator can be identified for a given system, the quantum wave functions,
apart from phase parts, can be expressed by eigenfunctions of this operator.
Consequently, solving the quantum system reduces to finding the
eigenfunctions of the invariant operator, while the required phase factors
can be determined with the help of the Schr\"{o}dinger equation.

Regarding coupling of potentials, as well as the system's time dependence,
the general form of the invariant operator is typically 
complicated.
Therefore, to solve its eigenvalue equation, more advanced mathematical
techniques, such as the unitary transformation method \cite{utm,17} and the
Nikiforov-Uvarov functional analysis (NUFA) \cite{21}, are necessary.
The NUFA method builds upon and integrates several approaches, including the
original Nikiforov-Uvarov (NU) method \cite{23}, the parametric NU method
\cite{24,18,19,20}, and elements of functional analysis \cite{25}. This enhanced
method enables one to obtain the solution of related equations, including those of
hypergeometric type, in a systematic and elegant manner, allowing the determination of
the eigenvalue spectrum and the corresponding
wave functions. We demonstrate that the wave functions derived through this
approach can be expressed in terms of the well-known Gauss hypergeometric
function. This may provide a significantly more tractable framework for
analyzing the system compared to existing solutions that are based on
biconfluent Heun functions.
Taking advantage of these strengths, we show that the newly obtained solutions
enable a thorough investigation of the system.
\newline
\newline
\textbf{2. Analysis of the System } \newline
\textbf{2.1. Basic formulations: Hamiltonian and invariant} \newline
{\it 2.1.1. Setup of the Hamiltonian} \\
We consider a quark-antiquark pair with a time-dependent effective reduced mass $\mu (t)$ in
spherical coordinates, governed by a non-central potential
$g(t)r^{2}-\frac{Z(t)}{r}$ where $g(t)$ and $Z(t)$ are
coefficients which depend on time. While $g(t)r^{2}$ represents a confining trap potential
with adjustable strength, $-\frac{Z(t)}{r}$ is a 
Coulomb perturbation
with the constraint $Z(t)>0$.
As the distance between quarks increases, the harmonic potential rises sharply,
effectively prohibiting their free separation within a bound state. This potential
acts as a simplified representation of the QCD confinement mechanism and dominates
at large separations.
Meanwhile, the Coulomb term represents the electromagnetic or color interaction mediated by the
color charge of one quark acting on the other, effectively serving as a short-range
attractive force between the quark and antiquark.

Based on the above description, the Hamiltonian that characterizes the motion of the pair of quarks is expressed
as:
\begin{equation}
H(t)=\frac{p^{2}}{2\mu (t)}+g(t)r^{2}-\frac{Z(t)}{r} .  \label{1}
\end{equation}
This Hamiltonian enables the calculation of the quantum mechanical spectrum
including wave functions and excitation properties, which is then
used to theoretically explain the mass and structure of mesons.
The momentum operator in spherical coordinates is represented as $%
p^{2}=p_{r}^{2}+\frac{L^{2}}{r^{2}}$, where $p_{r}=-i\hbar (\frac{\partial }{%
\partial r}+\frac{1}{r})$ and $L$ is the total angular momentum of which formula is given by
\begin{equation}
L^{2}=-\hbar ^{2}\bigg[\frac{1}{\sin ^{2}\theta }\frac{\partial ^{2}}{%
\partial \varphi ^{2}}+\frac{1}{\sin \theta }\frac{\partial }{\partial
\theta }\bigg(\sin \theta \frac{\partial }{\partial \theta }\bigg)\bigg].
\label{2}
\end{equation}%
To investigate dynamical features 
of the system, we must solve the Schr%
\"{o}dinger equation for the Hamiltonian in Eq. (\ref{1}). If we denote
system's state vector as $|\Psi (t)\rangle$, the corresponding Schr\"{o}dinger
equation is of the form
\begin{equation}
i\hbar \frac{\partial }{\partial t}|\Psi (t)\rangle =H(t)|\Psi (t)\rangle .
\label{3}
\end{equation}%
The time dependence of the Hamiltonian requires
a specialized approach to solving this equation, as will be presented
subsequently.
\\
{\it 2.1.2. Formulation of the invariant operator} \\
Because $H(t)$ in Eq. (\ref{1}) is represented in terms of time functions, directly solving Eq. (\ref{3})
can be very challenging. To overcome this difficulty, we
adopt the dynamical invariant method \cite{29}.
The core idea of this method is to treat the system in terms of a
non-trivial invariant operator $I(t)$ rather than $H(t)$ in Eq. (\ref{1}),
where $I(t)$ does not involve time derivative operators.
Moreover, $I(t)$ is closely related to the concept of invariants in classical
mechanics (e.g., constants of motion), thereby facilitating interpretation
from the perspective of quantum-classical correspondence.

By the definition
of the invariant operator, $I(t)$ satisfies the equation:
\begin{equation}
\frac{dI}{dt}=\frac{\partial I}{\partial t}+\frac{1}{i\hbar }[I,H]=0.
\label{4}
\end{equation}%
If such an invariant exists, it is possible to put the solution of the
Schr\"{o}dinger equation, Eq. (3), in terms of its eigenfunction $\Phi (\vec{r},t)$, such that
\begin{equation}
\Psi (\vec{r},t)=e^{i\alpha (t)}\Phi (\vec{r},t),  \label{5}
\end{equation}%
where $\alpha (t)$ is a phase function that can be obtained from
\begin{equation}
\hbar \frac{d\alpha (t)}{dt}=\left\langle \Phi \left\vert \bigg(i\hbar \frac{%
\partial }{\partial t}-H\bigg)\right\vert \Phi \right\rangle .  \label{6}
\end{equation}%
To find the expression of the invariant $I(t)$ utilizing Eq. (\ref{4}), we set
\begin{equation}
I(t)=A(t)p^{2}+B(t)\left( rp_{r}+p_{r}r\right) +C(t)r^{2}-\frac{D(t)}{r},
\label{7}
\end{equation}%
where $A(t)$, $B(t)$, $C(t)$, and $D(t)$ are time-dependent functions to
be determined. Then, a mathematical procedure with
this assumed formula of $I(t)$ after substituting Eqs. (\ref{1}) and (\ref{7}%
) into Eq. (\ref{4}) yields 
\begin{eqnarray}
A(t) &=&A_{0}\rho ^{2}(t),  \label{co1} \\
B(t) &=&-A_{0}\mu (t)\rho (t)\dot{\rho}(t),  \label{co2} \\
C(t) &=&A_{0}\bigg(\mu ^{2}(t)[\dot{\rho}(t)]^{2}+\frac{\Omega ^{2}}{4\rho
^{2}(t)}\bigg),  \label{co3} \\
D(t) &=&2A_{0}\mu (t)Z(t)\rho ^{2}(t),  \label{co4}
\end{eqnarray}%
where $A_{0}$ is an arbitrary real constant and $\rho (t)$ is a solution of
the following equation
\begin{equation}
\ddot{\rho}(t)+\frac{\dot{\mu}(t)}{\mu(t) }\dot{\rho}(t)+2\frac{g(t)}{\mu (t)}\rho(t) =%
\frac{\Omega ^{2}}{4\mu ^{2}(t)\rho ^{3}(t)}, \label{co4-1}
\end{equation}%
with $\Omega $ being a real constant. The above invariant
is valid under the condition that $Z(t)$ follows the relation
\begin{equation}
\dot{Z}(t)+\bigg(\frac{\dot{\mu}(t)}{\mu (t)}+\frac{\dot{\rho}(t)}{\rho (t)}%
\bigg)Z(t)=0.  \label{19-}
\end{equation}%
Equation (\ref{7}) with Eqs. (\ref{co1})-(\ref{co4}) constitutes the complete quadratic
invariant operator of the system.
This operator provides a robust analytical framework for the system, in which direct
solutions are difficult to obtain due to the complicated time evolution of the Hamiltonian.
It also enables a deeper understanding of dynamical properties of the system, including phase transitions and
topological changes, beyond merely providing a powerful methodology for obtaining solutions.
\\
{\it 2.1.3. Solving the eigenvalue equation} \\
Because the quantum wave functions of the system are represented in terms of
the eigenfunctions of $I(t)$, it is now necessary to evaluate its eigenvalue
equation. We begin by writing the eigenvalue equation of the invariant $I(t)$ as
\begin{equation}
I(t)\Phi _{n}(\vec{r},t)=\Lambda _{n}\Phi _{n}(\vec{r},t),  \label{11}
\end{equation}%
where $\Lambda _{n}$ are the eigenvalues and $\Phi _{n}(\vec{r},t)$ are
time-dependent eigenfunctions.
Given that $I(t)$, as defined in Eq. (\ref{7}) along with Eqs. (\ref{co1})-(\ref{co4}),
has a complicated form,
it is favorable to solve Eq. (\ref{11}) after we transform it mathematically into a simple
form. To do this, we consider a unitary transformation of the form
\begin{equation}
\Phi _{n}^{\prime }(\vec{r})=U(t)\Phi _{n}(\vec{r},t),  \label{12}
\end{equation}%
where $U(t)$ is a time-dependent unitary operator given by
\begin{equation}
U(t)=\exp \left[ \frac{i\ln \rho (t)}{2\hbar }\left( rp_{r}+p_{r}r\right) %
\right] \exp \left[ -i\frac{\mu (t)\dot{\rho}(t)}{2\hbar \rho (t)}r^{2}\right]
.  \label{13}
\end{equation}%
Then, if we write the transformed invariant operator as $I_0$, the eigenvalue equation in the transformed frame is
represented in terms of $\Phi _{n}^{\prime }(\vec{r})$, such that
\begin{equation}
UIU^{-1}\Phi _{n}^{\prime }(\vec{r})=I_{0}\Phi _{n}^{\prime }(\vec{r}%
)=\Lambda _{n}\Phi _{n}^{\prime }(\vec{r}).  \label{14}
\end{equation}%
Now, from a mathematical procedure using Eq. (\ref{13}), we obtain that
\begin{equation}
\left[ A_{0}p^{2}+A_{0}\frac{\Omega ^{2}}{4}r^{2}-D_{0}\frac{1}{r}\right]
\Phi _{nlm}^{\prime }(r,\theta ,\varphi )=\Lambda _{nl}\Phi _{nlm}^{\prime
}(r,\theta ,\varphi ),  \label{15}
\end{equation}%
where $D_{0} = 2A_{0}\mu (t)Z(t)\rho(t) $ while $n=0,1,2,\cdots $.
An infinite number of bound states may arise from the strong confinement of quarkonium,
which is primarily attributed to the harmonic term dominating at large distances.
By the way, the direct differentiation of $D_0$ with respect to $t$,
followed by the application of Eq. (\ref{19-}),
yields $d D_0/dt=0$. This means that $D_0$ is in fact a constant of motion,
as expected—since the operator on the left-hand side of Eq. (\ref{15}) is the transformed
version of the original invariant $I(t)$.
It is worth noting that Eq. (\ref{15}) is very simple compared to
the original eigenvalue equation represented in Eq. (\ref{11}).

The existence of a perturbation of Coulomb potential may make the problem much
more difficult.
In what follows, the eigenvalue equation, Eq. (\ref{15}), can be rewritten as
\begin{equation}
\left[
\begin{array}{c}
-\hbar ^{2}A_{0}\Big[\frac{1}{r^{2}}\frac{\partial }{\partial r}\left( r^{2}%
\frac{\partial }{\partial r}\right) +\frac{1}{r^{2}\sin ^{2}\theta }\frac{%
\partial ^{2}}{\partial \varphi ^{2}} \\
+\frac{1}{r^{2}\sin \theta }\frac{\partial }{\partial \theta }(\sin \theta
\frac{\partial }{\partial \theta })\Big]+A_{0}\frac{\Omega ^{2}}{4}r^{2}-%
\frac{D_{0}}{r}%
\end{array}%
\right] \Phi _{nlm}^{\prime }(r,\theta ,\varphi )=\Lambda _{nl}\Phi
_{nlm}^{\prime }(r,\theta ,\varphi ).  \label{16}
\end{equation}%
According to the invariant operator theory, the wave functions of the transformed system
are represented in terms of $\Phi _{nlm}^{\prime }(r,\theta ,\varphi )$.
Because this equation is independent of time, we can apply the method of
separation of variables in order to solve it. Considering this, we take
\begin{equation}
\Phi _{nlm}^{\prime }(r,\theta ,\varphi )=R_{nl}(r)Y_{lm}(\theta
,\varphi )=\frac{u_{nl}(r)Y_{lm}(\theta,\varphi )}{r},  \label{17}
\end{equation}%
where we have introduced $u_{nl}(r) = r R_{nl}(r)$, whereas $Y_{lm}(\theta ,\varphi )$ are the spherical harmonics.
The angular part of the eigenfunctions takes the form
\begin{equation}
Y_{lm}(\theta ,\varphi )= N_{lm}e^{im\varphi }P_{l}^{m}(\cos
\theta ),  \label{18}
\end{equation}%
where $P_{l}^{m}(x)$ are the associated Legendre polynomials,
while the spherical normalization constants are
$N_{lm}=(-1)^m \{ [(2l+1)/(4\pi)](l-m)!/(l+m)! \}^{1/2}$.
The allowed orbital and magnetic quantum numbers are $l=0,1,2,\cdots$ and $m=-l,-l+1,\cdots,l$, respectively.

Some algebraic manipulation, after substituting Eq. (\ref{17}) into Eq. (\ref{16}), leads
to the following radial equation:
\begin{equation}
\frac{d^{2}u_{nl}(r)}{dr^{2}}+\left( \frac{\Lambda _{nl}}{\hbar ^{2}A_{0}}-%
\frac{\Omega ^{2}}{4\hbar ^{2}}r^{2}+\frac{D_{0}}{\hbar ^{2}A_{0}}\frac{1}{r}%
-\frac{l(l+1)}{r^{2}}\right) u_{nl}(r)=0.  \label{19}
\end{equation}%
The equation is analytically intractable unless $\Omega = 0$ owing to the
coexistence of $r^{2}$ and $-\frac{1}{r^{2}}$ terms. To resolve this difficulty, we employ the
Greene-Aldrich approximation scheme of the form \cite{22,gaa1,gaa2,gaa3}
\begin{equation}
\frac{1}{r^{2}} \simeq \frac{\delta ^{2}}{\left( 1-e^{-\delta r}\right) ^{2}}%
;~~~~~\frac{1}{r} \simeq \frac{\delta }{1-e^{-\delta r}},  \label{20}
\end{equation}%
together with a coordinate transformation $y=e^{-\delta r}$.
This type of approximation is commonly used to describe particles governed by a screened
potential, where the parameter $\delta$ is typically chosen to be equal to the screening mass \cite{21,22}.
Though the potential in our system is not screened, choosing $\delta$ to be similar to the typical
screening mass still provides a good approximation. For example, in quarkonium models, $\delta$
is usually taken to be in the range of approximately 200 to 900 MeV \cite{sm1,sm2}.
Thus, selecting $\delta$
within this range is a reasonable choice.

Then, by rewriting the notation as $u_{nl}(r)\rightarrow U_{nl}(y)$, the differential equation,
Eq. (\ref{19}), becomes
\begin{equation}
\frac{d^{2}U_{nl}(y)}{dy^{2}}+\frac{\left( 1-y\right) }{y(1-y)}\frac{%
dU_{nl}(y)}{dy}+\frac{1}{y^{2}(1-y)^{2}}\left[
\begin{array}{c}
-\left( \varepsilon _{nl}-6P\right) y^{2} \\
+\left( 2\varepsilon _{nl}-4P+Q\right) y \\
-\left( \varepsilon _{nl}-P+Q+\gamma \right)%
\end{array}%
\right] U_{nl}(y)=0,  \label{21}
\end{equation}%
where
\begin{equation}
\varepsilon _{nl}=\frac{-\Lambda _{nl}}{\hbar ^{2}A_{0}\delta ^{2}},~~~P=-%
\frac{\Omega ^{2}}{4\hbar ^{2}\delta ^{4}},~~~Q=-\frac{D_{0}}{\hbar
^{2}A_{0}\delta },~~~\gamma =l(l+1).  \label{22}
\end{equation}%
In the representation of Eq. (\ref{21}), we neglected
higher order terms corresponding to $y^{3}$ and $y^{4}$.
To solve this equation, we use the NUFA method \cite{21} outlined in the next subsection.
\newline
\textbf{2.2. The NUFA method} \newline
A second order differential equation of hypergeometric type can be
solved in a simple and elegant way by using the so-called NUFA.
This method is an improvement
over the parametric NU method, which is relatively cumbersome due to the
need to find the square of the polynomials and other conditions required.
With the NUFA method, many diffierential equations, including our wave equation, can be properly tackled,
allowing for the identification of singularities and the determination of the
eigenvalue spectrum.
As a general type of second order differential
equation, which encompasses Eq. (\ref{21}), we consider the following equation \cite{24}
\begin{equation}
\left[ \frac{d^{2}}{ds^{2}}+\frac{\alpha _{1}-\alpha _{2}s}{s(1-\alpha _{3}s)%
}\frac{d}{ds}+\frac{-\zeta _{1}s^{2}+\zeta _{2}s-\zeta _{3}}{s^{2}(1-\alpha
_{3}s)^{2}}\right] \phi (s)=0,  \label{24}
\end{equation}%
where $\alpha _{i}$$(i=1,2,3)$ and $\zeta _{i}$ are parameters appropriately
chosen depending on a given system. Because this
equation has two singularities at $%
s\rightarrow 0$ and $s\rightarrow 1/\alpha _{3}$, it is possible to take the
solution 
in the form
\begin{equation}
\phi (s)=s^{\lambda }(1-\alpha _{3}s)^{\upsilon }f(s).  \label{25}
\end{equation}%
By substituting this formula into Eq. (\ref{24}) along with the following
choice
\begin{eqnarray}
\lambda &=&\frac{1}{2}\{ 1-\alpha _{1}\pm [\left( 1-\alpha
_{1}\right) ^{2}+4\zeta _{3}]^{1/2}\} ,  \label{26}
\\
\upsilon &=&\frac{1}{2\alpha _{3}} \{ \alpha _{3}+\alpha _{1}\alpha
_{3}-\alpha _{2}\pm [\left( \alpha _{3}+\alpha _{1}\alpha _{3}-\alpha
_{2}\right) ^{2} \nonumber \\
& &+4\alpha_3 (\zeta _{3}\alpha
_{3}-\zeta _{2}+\zeta _{1}/\alpha _{3})]^{1/2}\} ,  \label{27}
\end{eqnarray}%
we have \cite{21}
\begin{equation}
s(1-\alpha _{3}s)\frac{d^{2}f(s)}{ds^{2}}+\kappa \frac{df(s)}{ds}
-\alpha _{3} \kappa_+ \kappa_- f(s)=0,  \label{28}
\end{equation}%
where
\ba
\kappa &=&  \alpha _{1}+2\lambda -(2\lambda \alpha _{3}+2\upsilon \alpha _{3}+\alpha _{2})s, \\
\kappa_\pm &=& \lambda +\upsilon +\frac{1}{2}\left( \frac{\alpha _{2}}{%
\alpha _{3}}-1\right) \pm \bigg[ \frac{1}{4}\left( \frac{\alpha _{2}}{\alpha _{3}%
}-1\right) ^{2}+\frac{\zeta _{1}}{\alpha _{3}^{2}} \bigg]^{1/2}.
\ea
Thus, the equation is reduced to a tractable form.

We now examine a simple case where $\alpha_3=1$, which meets the equation
associated with our considered system.
Under the condition that the principal quantum number takes the form
\begin{equation}
n=- \kappa_- ,
\label{29}
\end{equation}%
the function $f(s)$ is represented by the Gauss hypergeometric function as \cite{21}
\begin{eqnarray}
f (s) = {_{2}F_{1}}(a,b;c;s),  \label{31}
\end{eqnarray}%
where
$a=\kappa_- $, $b=\kappa_+ $, and $c=\alpha _{1}+2\lambda $.
The mathematical formula of this hypergeometric function is as follows: 
 $_{2}F_{1}(a,b;c;s)=\sum_{k=0}^{\infty }\frac{\left( a\right)
_{k}\left( b\right) _{k}}{\left( c\right) _{k}k!}s^{k}$, where $\left(
a\right) _{k}=a(a+1)(a+2)\cdots (a+k-1)$ for $k\geq 1$ and $\left( a\right)
_{0}=1$.

The Gauss hypergeometric function $_{2}F_{1}(a,b;c;s)$ serves as a fundamental and unifying special
function that encompasses many other functions as special or limiting
cases \cite{2f1,2f2,2f3,NIST}.
As defined in Eq. (\ref{31}) with Eq. (\ref{29}), the function $_{2}F_{1}(a,b;c;s)$ terminates and
reduces to a polynomial when $a$ is zero or a negative integer, a case that is
often associated with Laguerre polynomials.
In particular, $_{2}F_{1}(a,b;c;s)$
is a generalization of Kummer’s confluent hypergeometric function $_{1}F_{1}(a;c;s)$.
Hence, in the limiting case where $b \rightarrow \infty$,
$_{2}F_{1}(a,b;c;s)$ converges to $_{1}F_{1}(a;c;s)$.
This convergence is rooted in the fact that, while
${}_2F_1$ has two regular singular points, sending
one of them to infinity (a process known as confluence)
effectively merges the singularities, giving rise to ${}_1F_1$ \cite{2f3}.
Additionally, the Jacobi polynomials can also be expressed in terms of the Gauss
hypergeometric function (for a typical representation, see formula 15.9.1 of Ref. \cite{NIST}).
The ${}_2F_1$ function thus plays a major
role in constructing analytical solutions and enables
rigorous, systematic analysis
across a broad spectrum of scientific disciplines beyond mathematics and physics.
The method presented in this subsection, which naturally leads to appearing
the ${}_2F_1$ function as a solution, is applicable to
the analysis of a wide class of physical systems
governed by one or more interaction potentials.
It is, of course, well-suited for handling Eq. (\ref{21}).
\newline
\textbf{2.3. Wave functions} \newline
{\it 2.3.1. Eigenfunctions of the invariant operator} \\
We now proceed to derive the wave functions of the system.
For clarity and better reader understanding, we briefly summarize our strategy at this point.
Because, in our time-dependent case, the wave functions are described by the eigenfunctions
of $I(t)$ (instead of $H(t)$), we need to determine the eigenfunctions
$\Phi_n$ of Eq. (\ref{11}).
However, directly solving for $\Phi_n$ is likely
to be a formidable task. To circumvent this difficulty, we first derive
$\Phi_n'$ from Eq. (\ref{14}), and then
reconstruct the desired eigenfunctions $\Phi_n$ by inversely transforming
the obtained $\Phi_n'$, utilizing the inverse relation of 
Eq. (\ref{12}) (for the inverse relation, see Eq. (\ref{47}), which appears later).
Once $\Phi_n$ are obtained, the wave functions $\Psi_n$ can be expressed in
terms of them accordingly.

To obtain $\Phi_n'$, we need to solve Eq. (\ref{21}) by setting the associated functions, in accordance
with Eq. (\ref{25}), as
\begin{equation}
U_{nl}(y)=N_{nl}y^{\lambda }(1-y)^{\upsilon }f_{nl}(y),  \label{}
\end{equation}%
where $N_{nl}$ are normalization constants for the final wave functions.
We applied $\alpha _{3}=1$ in this equation.
In addition, from the comparison of Eq. (\ref{21}) with the NUFA equation (\ref{24}),
we acquire the following parameter relations:
\begin{eqnarray}
&&~\alpha _{1}=\alpha _{2}=1,\text{ \ \ \ \ \ \ \ \ \ \ \ \ }\zeta
_{1}=\varepsilon _{nl}-6P,~~~~~~  \notag \\
&&\zeta _{2}=2\varepsilon _{nl}-4P+Q,\text{ \ \ \ }\zeta _{3}=\varepsilon
_{nl}-P+Q+\gamma .~~~~~~   \label{39m}
\end{eqnarray}%
Furthermore, the singularity exponents reduce to
\begin{equation}%
\lambda =\sqrt{\varepsilon _{nl}-P+Q+\gamma },\text{ \ \ }\upsilon =\frac{1}{%
2}+\sqrt{\left( l+\frac{1}{2}\right) ^{2}-3P}.~~~~~~  \label{39}
\end{equation}%
Although the original NUFA representations of $\lambda$ and $\upsilon$,
given in Eqs. (\ref{26}) and (\ref{27}), include both positive and
negative signs ($\pm$) from a purely mathematical perspective,
we choose to retain only the positive sign in this case as can be seen from Eq. (\ref{39}).
This selection is due to the fundamental physical constraint, which is that the wave functions must remain finite throughout
the entire range of $r$ and must vanish as $r \rightarrow 0$ and $r \rightarrow \infty$.

The expansion of Eq. (\ref{29}) under the first condition in Eq. (\ref{39m}) yields
\begin{equation}
n+\lambda+\upsilon = \sqrt{\zeta_1}.
\end{equation}
By squaring both sides of this equation and using the second relation from Eq. (\ref{39m})
together with the formula of $\lambda$ given in Eq. (\ref{39}), the equation associated with the
eigenvalues can be obtained as
\begin{equation}
2[\varepsilon _{nl}-P+Q+\gamma ]^{1/2}(n+\upsilon )+(n+\upsilon
)^{2}+5P+Q+\gamma =0.  \label{40}
\end{equation}%
It is possible to rearrange this equation into a formula for $\varepsilon _{nl}$ by first
isolating the square root term on the left-hand side, shifting all
other terms to the right-hand side, and then squaring both sides.
Substituting the expression for $\upsilon$ from Eq. (\ref{39}) into the resulting equation,
and using Eq. (\ref{22}), we finally obtain the eigenvalues:
\begin{eqnarray}
\Lambda _{nl} &=&l(l+1)\hbar ^{2}A_{0}\delta ^{2}+\frac{A_{0}\Omega ^{2}}{%
4\delta ^{2}}-D_{0}\delta   \notag \\
&&-\frac{\hbar ^{2}A_{0}\delta ^{2}}{4}\left[ \frac{l(l+1)-\frac{5\Omega ^{2}%
}{4\hbar ^{2}\delta ^{4}}-\frac{D_{0}}{\hbar ^{2}A_{0}\delta }+\left(
n+\upsilon \right) ^{2}}{n+\upsilon }\right] ^{2}.  \label{41}
\end{eqnarray}%
In addition, by considering Eq. (\ref{25}) with Eqs. (\ref{26}), (\ref{27}), and
(\ref{31}), the corresponding radial wave functions can
be derived in terms of $\Lambda_{nl}$ to be
\begin{eqnarray}
U_{nl}(y) &=&N_{nl}y^{\sqrt{-\frac{\Lambda _{nl}}{\hbar ^{2}A_{0}\delta ^{2}}%
+\frac{\Omega ^{2}}{4\hbar ^{2}\delta ^{4}}-\frac{D_{0}}{\hbar
^{2}A_{0}\delta }+l(l+1)}}  \notag \\
&&\times {(1-y)^{\frac{1}{2}+\sqrt{\left( l+\frac{1}{2}\right) ^{2}+\frac{%
3\Omega ^{2}}{4\hbar ^{2}\delta ^{4}}}}}{_{2}F_{1}}(\bar{a},\bar{b};\bar{c}%
;y),  \label{42}
\end{eqnarray}%
where the parameters are defined as
\begin{eqnarray}
&&\bar{a}=\lambda +\upsilon -\sqrt{\varepsilon _{nl}-6P},  \label{43} \\
&&\bar{b}=\lambda +\upsilon +\sqrt{\varepsilon _{nl}-6P},  \label{44} \\
&&\bar{c}=1+2\lambda .  \label{45}
\end{eqnarray}%
From a mathematical standpoint, it is possible to define the parameter $\bar{a}$ interchangeably
with $\bar{b}$. However, by choosing $\bar{a}$ as in Eq. (\ref{43}), we allow the possibility for $\bar{a}$
to take on negative values. When $\bar{a}$ is a non-positive integer (i.e., a negative integer
or zero), the Gauss hypergeometric function in Eq. (\ref{42})
terminates at the ($n+1$)-th term and reduces to a polynomial.
Now, through the use of Eq. (\ref{17}) with Eq. (\ref{42}),
the full eigenfunctions of the invariant operator $I_{0}$ are represented in the form
\begin{eqnarray}
\Phi _{nlm}^{\prime }(r,\theta ,\varphi ) &=&N_{nl}\frac{1}{r}e^{-\left(
\sqrt{-\frac{\Lambda _{nl}}{\hbar ^{2}A_{0}\delta ^{2}}+\frac{\Omega ^{2}}{%
4\hbar ^{2}\delta ^{4}}-\frac{D_{0}}{\hbar ^{2}A_{0}\delta }+l(l+1)}\right)
\delta r}  \notag \\
&&\times \left( 1-e^{-\delta r}\right) ^{\frac{1}{2}+\sqrt{\left( l+\frac{1}{%
2}\right) ^{2}+\frac{3\Omega ^{2}}{4\hbar ^{2}\delta ^{4}}}}{_{2}F_{1}}(\bar{%
a},\bar{b};\bar{c};e^{-\delta r})Y_{lm}(\theta ,\varphi ).  \label{46}
\end{eqnarray}%

The complete
eigenfunctions of the invariant $I$ that belongs to the original system,
i.e., the untransformed system, are given from the relation
\begin{equation}
\Phi _{nlm}(\vec{r},t)=U^{-1}(t)\Phi _{nlm}^{\prime }(\vec{r}).  \label{47}
\end{equation}%
The evaluation of this using Eqs. (\ref{13}) and (\ref{46}) results in 
\begin{eqnarray}
\Phi _{nlm}(r,\theta ,\varphi ,t) &=&N_{nl} \frac{1 }{\sqrt{\rho(t)}r}
\exp \left( \frac{i\mu (t)\dot{\rho}(t)}{2\hbar \rho (t)}%
r^{2}\right)  \notag  \\
&&\times  e^{-\left( \sqrt{-\frac{\Lambda _{nl}%
}{\hbar ^{2}A_{0}\delta ^{2}}+\frac{\Omega ^{2}}{4\hbar ^{2}\delta ^{4}}-%
\frac{D_{0}}{\hbar ^{2}A_{0}\delta }+l(l+1)}\right) \delta \frac{r}{\rho(t) }}
\notag \\
&&\times \left( 1-e^{-\delta \frac{r}{\rho(t) }}\right) ^{\frac{1}{2}+\sqrt{%
\left( l+\frac{1}{2}\right) ^{2}+\frac{3\Omega ^{2}}{4\hbar ^{2}\delta ^{4}}}%
}{_{2}F_{1}}(\bar{a},\bar{b};\bar{c};e^{-\delta \frac{r}{\rho(t) }%
})Y_{lm}(\theta ,\varphi ).  \label{48}
\end{eqnarray}%
Meanwhile, $N_{nl}$ are derived by normalizing these resultant eigenfunctions.
The normalization procedure yields
(see Appendix A for detailed evaluation)
\begin{equation}
N_{nl}=\sqrt{\delta} \f{\Gamma (2(\upsilon+\lambda)+n)}{n!\Gamma(2\lambda+1)} \Bigg[
\sum_{k=0}^n\sum_{k'=0}^n (-1)^{k+k'} \mathcal{N}_{nl}(k,k') \Bigg]^{-1/2}, \label{normal}
\end{equation}
where
\ba
\mathcal{N}_{nl}(k,k') &=& \f{\Gamma(2(\upsilon+\lambda)+n+k)
\Gamma(2(\upsilon+\lambda)+n+k')}{k!k'!(n-k)!(n-k')!\Gamma(2\lambda+k+1)\Gamma(2\lambda+k'+1)}
\nonumber \\
& &\times{\mathrm B}(k+k'+2\lambda,2\upsilon+1),
\ea
whereas ${\mathrm B}(a_1,a_2)$ denotes the beta function \cite{beta}.
The eigenfunctions in Eq. (\ref{48}) are essentially related to the Schr\"{o}dinger solutions
of the system.
We demonstrate how to construct the full quantum solutions in terms of these eigenfunctions
in the following subsection.
\newline
{\it 2.3.2. Quantum phases and the Schr\"{o}dinger solutions} \\
According to Lewis-Riesenfeld theory \cite{29}, the wave functions
of our time-dependent system
are represented in terms of the eigenfunctions $\Phi_{nlm}(\vec{r},t)$ of the
invariant operator, which are given in Eq. (\ref{48}).
However, these eigenfunctions alone do not fully define the time-dependent
wave functions. Another required factor is the phase $\alpha (t)$
associated with each quantum state.
We begin from Eq. (\ref{6}) to derive this factor.
Based on the unitary relation given in Eq. (\ref{47}), Eq. (\ref{6}) can be
expressed in terms of the transformed eigenstate $\vert \Phi^\prime \rangle$
instead of $\vert \Phi \rangle$, such that
\begin{equation}
\hbar \frac{d\alpha (t)}{dt}=\left\langle \Phi ^{\prime }\left\vert -\frac{%
I_{0}}{2A_{0}\mu (t)\rho ^{2}(t)}\right\vert \Phi ^{\prime }\right\rangle .
\label{50}
\end{equation}%
Then, a minor evaluation using Eq. (\ref{15}) yields the phases in the form
\begin{equation}
\alpha _{nl}(t)=-\f{\Lambda _{nl}}{2A_{0}\hbar}\int_{0}^{t}\frac{1}{\mu (t^{\prime
})\rho ^{2}(t^{\prime })}dt^{\prime },  \label{51}
\end{equation}%
where $\Lambda _{nl}$ are given by Eq. (\ref{41}).
These phases emerge as a result of the time evolution of the the wave functions
and can produce observable consequences, such as interference patterns, through
relative phase differences between quantum states.
Both the geometrical phase and the conventional
dynamical phase contribute to the total phase. While the dynamical
phase results from the cumulation of energy over time, the geometric
phase originates from the intrinsic geometry of the path traced by
the state vector during evolution, often under adiabatic conditions.
The significance of the geometrical phase lies in the fact that it is not
a mathematical artifact but a physically meaningful, gauge-invariant quantity that
can influence the actual quantum state of the system.
It not only takes place in this system, but appears ubiquitously across
various physical systems as well, and can be
measured via interference experiments.
Unlike the dynamical phase, which depends on the eigenvalues of the Hamiltonian,
the geometrical phase depends on the trajectory of the eigenvectors themselves.
As such, it plays a critical role in the associated phenomena such as energy level splitting and
the orbital behavior of quasiparticles, making it a subject of continued theoretical
and experimental interest. In particular, in systems with irregular and nonlinear
interactions, such as quark-antiquark systems, it reflects the underlying symmetries,
binding mechanisms, and topological features of the system \cite{gp1,gp2}.
Accordingly, the geometrical phase constitutes a fundamental component
in the evolution of the system, linking theoretical descriptions
to experimental observations.

The solutions of
the original Schr\"{o}dinger equation, Eq. (\ref{3}), corresponding to the
Hamiltonian in Eq. (\ref{1}), can now be written in configuration space as
\begin{equation}
\Psi _{nlm}(r,\theta ,\varphi ,t)=e^{i\alpha _{nl}(t)}\Phi _{nlm}(r,\theta
,\varphi ,t),  \label{52}
\end{equation}%
where the explicit forms of $\alpha _{nl}(t)$ and $\Phi _{nlm}(r,\theta
,\varphi ,t)$ are given in Eqs. (\ref{51}) and (\ref{48}), respectively.
The resulting wave functions, Eq. (\ref{52}), together with
$\Lambda_{nl}$ in Eq. (\ref{41}), serve as fundamental solutions for the
quark-antiquark systems under consideration.
The obtained wave functions enable us to compute spatial distributions (densities),
transition probabilities, spin-dependent interactions, etc.
Consequently, they provide a
foundational framework for analyzing these systems, advancing our perspective
on the interaction mechanisms between quarks and antiquarks,
particularly in the context of quantum effects.
The spatial structure of the wave functions enables estimation of key physical properties,
such as the size of the particle (e.g., root-mean-square radius) and the
binding strength of the bound state, which are critical for characterizing the complexity of strong
interaction systems \cite{wf1,wf2}. These wave functions thus serve as important tools for
bridging theoretical predictions and experimental results, helping reveal the internal
structure, confinement dynamics, and symmetry properties of the quark-antiquark pair.
\newline
\newline
\textbf{3. Conclusion} \newline
In this study, we investigated the interaction between a quark and an antiquark by modeling
its mechanical behavior using a non-stationary 3D
harmonic oscillator coupled with a Coulomb potential.
To solve the associated Schr\"{o}dinger equation, we employed the
recently refined NUFA method, together with
the invariant operator method and the unitary transformation approach.
The NUFA method, based on a novel conceptual framework,
allowed us to 
bypass the complex mathematical manipulations commonly encountered in
other techniques. As a result, we obtained the wave functions of the system in closed form,
expressed in terms of the Gauss hypergeometric function.
These solutions offer advantages over previous ones that employed biconfluent
Heun functions, as Gauss hypergeometric functions are among the most thoroughly
characterized and best understood special functions in mathematics.

Our solutions are essential for analyzing quark-antiquark pairs,
which form the core of mesons and thus play
a central role in the strong interaction.
Although their dynamics are governed by QCD, the theory exhibits intricate
behavior that significantly deviates from classical, non-relativistic mechanics,
encompassing numerous unresolved challenges.
These include
the incomplete theoretical understanding of quark confinement and
hadronization mechanisms, the precise determination of quark masses and meson spectra,
and the possible existence of exotic hadrons featuring internal structure beyond conventional
quark-antiquark configurations \cite{eh1,eh2}.
Additional open questions involve the quantitative description of the quark condensate mechanism,
the strong CP problem and the role of the $\theta$-term, as well as the difficulty in accurately
predicting the spectral properties and decay modes of heavy mesons such as the $J/\psi$ \cite{np3,ee1}.
The analytical solutions developed in this study may provide
insights toward addressing these longstanding issues.

\appendix

\section{Derivation of the Normalization Constants}
Although normalization is formally defined for the full wavefunctions,
the phase factors do not affect the outcome. Therefore, in the normalization
process, it suffices to consider only equation (\ref{48}), which corresponds
to the eigenfunctions.
The normalization condition associated with Eq. (\ref{48}) is given by
\begin{equation}
\int_0^{2\pi} \int_0^\pi \int_0^\infty \left\vert \Phi_{nlm} (r,\theta,\varphi,t) \right\vert ^{2}r^{2}
\sin \theta dr d\theta d\varphi =1. \label{AP1}
\end{equation}
Because the angular part $Y_{lm} (\theta,\varphi)$, defined in Eq. (\ref{18}), is already normalized, we only need to
manage the radial part in this evaluation.
Considering the radial part of Eq. (\ref{48}), Eq. (\ref{AP1}) can be rewritten as
\begin{equation}
\frac{\left\vert N_{nl}\right\vert ^{2}}{\rho(t) }\int_{0}^{\infty
}e^{-2\lambda\delta \frac{r}{\rho (t)}}\left( 1-e^{-\delta \frac{r}{%
\rho (t)}}\right) ^{2\upsilon}\left[ {_{2}F_{1}}(\bar{a},\bar{b};\bar{%
c};e^{-\delta \frac{r}{\rho (t)}})\right] ^{2}dr =1.  \label{AP2}
\end{equation}
To simplify the integral, we introduce the substitution
\begin{equation}
z=e^{-\delta \frac{r}{\rho (t)}},
\end{equation}
and consider the fact that $\bar{a}$ and $\bar{b}$ in this case can be expressed in terms of $n$ as
\begin{equation}
\bar{a}=-n,~~~~~~~   \bar{b}=2(\upsilon+\lambda)+n,
\end{equation}
along with Eq. (\ref{45}) as the formula of $\bar{c}$. Then, Eq. (\ref{AP2}) becomes
\begin{equation}
\frac{\left\vert N_{nl}\right\vert ^{2}}{\delta }g_{\rm int}=1, \label{AP5}
\end{equation}
where $g_{\rm int}$ is the integral part:
\begin{equation}
g_{\rm int}=\int_{0}^{1}z^{2\lambda-1} \left( 1-z\right) ^{2\upsilon}\left[
{_{2}F_{1}}(-n,2(\upsilon+\lambda)+n;2\lambda+1;z)\right]
^{2}dz . \label{AP6}
\end{equation}
To evaluate the integration $g_{\rm int}$, we use the expansion
formula for the hypergeometric function (see formula 15.2.4 of Ref. \cite{NIST})
\begin{equation}
{_{2}F_{1}}(-n,b;c;s)=\sum_{k=0}^n (-1)^k \binom{n}{k} \f{(b)_k}{(c)_k}s^k,
\end{equation}
which holds when $n=0,1,2,\cdots$ and $c \neq 0,-1,-2,\cdots$.
This relation also holds when $c=-n-j$, where $j=0,1,2,\cdots$.
Substituting this series expansion into Eq. (\ref{AP6}), we obtain
\begin{eqnarray}
g_{\rm int}&=&\sum_{k=0}^n\sum_{k'=0}^n(-1)^{k+k'}\binom{n}{k} \binom{n}{k'}
\f{(2(\upsilon+\lambda)+n)_k (2(\upsilon+\lambda)+n)_{k'}}{(2\lambda+1)_k(2\lambda+1)_{k'}}
\nonumber \\
& &\times \int_{0}^{1}z^{k+k'+2\lambda-1} \left( 1-z\right) ^{2\upsilon}dz . \label{AP8}
\end{eqnarray}
The integral in the above equation is the Beta function, which is defined as \cite{beta}
\be
\mathrm{B}(a_1,a_2)= \int_0^1 s^{a_1-1} (1-s)^{a_2-1} ds,
\ee
under the conditions ${\rm Re}a_1>0$ and ${\rm Re}a_2>0$.
The two shape parameters of this function, in our case, are $a_1=k+k'+2\lambda$ and $a_2=2\upsilon+1$.
Thus, considering that the Pochhammer number is given by $(a)_k=\Gamma(a+k)/\Gamma(a)$,
where $a \neq 0,-1,-2,\cdots$,
a rearrangement of Eq. (\ref{AP5}) with Eq. (\ref{AP8}) leads directly to the expression
for the normalization constants $N_{nl}$,
as presented in Eq. (\ref{normal}) of the text.


\end{document}